\documentclass[apj]{emulateapj}
\usepackage{amsmath}
\usepackage{graphicx}
\usepackage{float}

\newcommand{\DDir}{\relax{D\kern-.7em{/}}}

\newcommand{\be}{\begin{equation}}
\newcommand{\ee}{\end{equation}}
\newcommand{\bea}{\begin{equation*}}
\newcommand{\eea}{\end{equation*}}

\newcommand{\nin}{\relax{\in\kern-.8em{/}}}

\newcommand{\sig}{\sigma}

\begin{document}
\newcommand{\vv}{\textrm{v}}
\title{Luminosity Function Suggests Up to 100 White Dwarfs within $20\,\rm pc$  may be hiding in multiple systems} 
\author{Boaz Katz\altaffilmark{1,2}, Subo Dong\altaffilmark{3},  Doron Kushnir\altaffilmark{1}}
 \altaffiltext{1}{Institute for Advanced Study, Einstein Drive, Princeton, New Jersey, 08540, USA} \altaffiltext{2}{John N.\ Bahcall Fellow} \altaffiltext{3}{Kavli Institute for Astronomy and Astrophysics, Peking University,Yi He Yuan Road 5, Hai Dian District, Beijing 100871, China}

\begin{abstract}
We examine the luminosity function of white dwarfs (WDs) in the local  ``complete'' WD sample ($d<20\,\rm pc$) of \citet{Holberg08}. We find that the fraction of bright and young WDs is anomalously high among the WDs detected in multiple systems with main sequence (MS) companions compared to that of the single WDs and theoretical expectations. This indicates a significant observation bias against finding relatively faint WDs in multiple systems. At the bright end ($M_V<11.5$), the amount of WDs with MS companions is approximately equal to that of single white dwarfs, indicating that $\gtrsim 50\%$ of WDs have MS companions, consistent with the high multiplicity fraction of early-type MS stars. If true, a significant fraction of WDs in multiple systems within $20\,\rm pc$ may have not been detected yet, and the number density of WDs in the solar neighborhood and elsewhere may be up to twice as much as presently believed.    
\end{abstract}

\section{Introduction}
It was recently noted by \citet{Ferrario12} that there is tension between the high multiplicity fraction ($>50\%$) of main sequence (MS) stars more massive than the Sun \citep[e.g.][]{Raghavan10}  and the low multiplicity fraction ($\sim 30\%$) of white dwarfs (WDs) which are the end states of these stars \citep{Holberg09}. The WD multiplicity is inferred from the sample of WDs within $20~\rm pc$ from the Sun, which is believed to be $80\%$ complete \citep{Holberg08}.  \citet{Ferrario12} suggested that the main reason for this discrepancy is that many WDs in multiple systems have not been detected simply because they are too faint relative to their companions. Here we present direct observational evidence suggesting that this is likely the case by studying the luminosity function of the detected WDs.

\section{Evidence that White Dwarfs with MS companions are missing in samples of nearby stars}
It is instructive to first examine the WDs among the closest 100 stars \footnote{http://www.chara.gsu.edu/RECONS/TOP100.posted.htm}. In this sample, there are 4 WDs in multiple systems and 4 single WDs. The former ones have absolute visual magnitudes $M_{V} = 11.0, 11.3, 13.0, 13.7$, which are systematically brighter than the latter with $M_{V} = 13.2,14.2,15.3,15.4$. This pattern persists in the larger sample of nearby WDs within $20~\rm pc$ \citep{Holberg08}. The $M_{V}$ distributions in the \citet{Holberg08} sample are shown in Figure 1 for single WDs (92 WDs, blue line) and WDs with MS companions (24 WDs, red line), respectively. While the amount of bright WDs ($M_V<11.5$) is similar for these two sets, there are significantly fewer fainter WDs among the MS-WD binaries as compared to the single WDs. 

To a good approximation, the luminosity of a WD reflects its age. To illustrate this, the relation between luminosity and age estimates for the subsample of WDs in \citet{Holberg08} for which sufficient WD parameters are reported (mass, $M$, effective temperature, $T_{\rm eff}$ and surface gravity, $\log g$) are shown in Figure 2 (blue and red points for single WDs and WDs in WD-MS binaries respectively). The WD age is estimated according to \citet{Wood95} using the inferred bolometric luminosity $L_{\rm bol}=4\pi R^2\sig_BT_{\rm eff}^2$, where $R^2=GM/g$.  The ages at a given luminosity are insensitive to the mass of the WDs, changing by about $30\%$ across the available WD mass range of $0.5 M_{\odot}<M<0.9 M_{\odot}$ \citep[e.g.,][]{Renedo10}. While the model calculated by \citet{Wood95} is for hydrogen dominated atmospheres applicable to the majority of the WDs in the sample, the cooling sequences (as a function of luminosity) differ by tens of percents for helium dominated atmospheres \cite[e.g.][]{Hansen99}. These differences are ignored in our analysis.   

\begin{figure}[h!]
\centering
\includegraphics[width=0.5\textwidth]{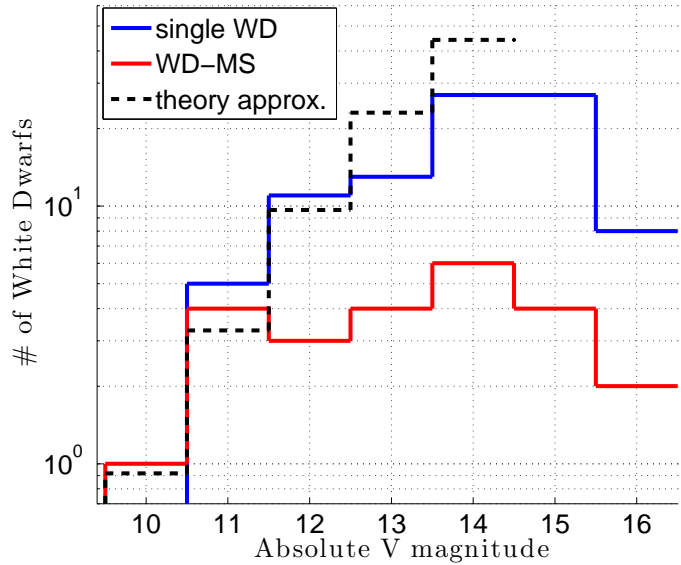}
\caption{Absolute visual magnitude distribution of WDs closer than $20\,\rm pc$ from \citet{Holberg08}. The distribution of isolated WDs (solid blue) is peaked at the faint brightness.  The distribution of WDs with MS companions (solid red) is roughly uniform. The lack of faint WDs with MS companions is likely an observational bias due to the difficulty in detecting close companions with a large contrast. An approximate expected distribution is shown for illustration (dashed black) assuming a constant rate of WD formation in the last $3$ Gyr and the relation between absolute visual magnitude and cooling age from Figure \ref{fig:Age}, based on the WD cooling model of \citet{Wood95}.
\label{fig:Mv}}
\end{figure}

\begin{figure}
\centering
\includegraphics[width=0.5\textwidth]{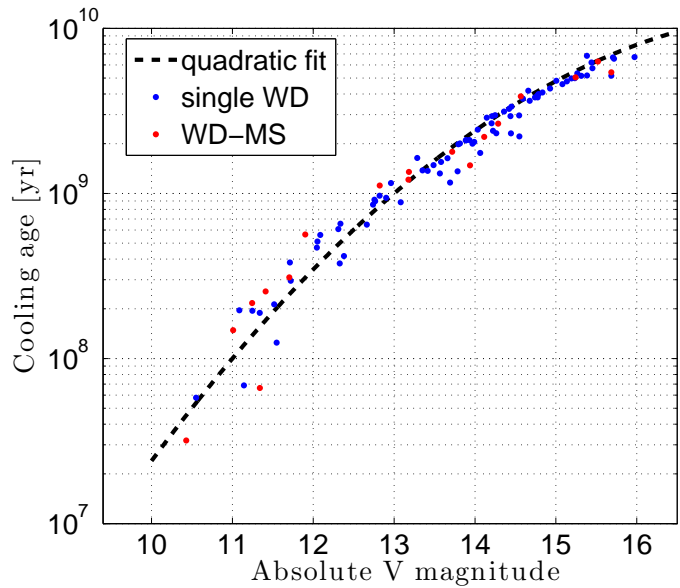}
\caption{Relation between absolute visual magnitude and cooling age. The cooling ages for the subsample of WDs in \citep{Holberg08} that have reported $T_{\rm eff},\log g$ and mass are estimated based on their deduced bolometric luminosity using the cooling model of \citet{Wood95}. Isolated WDs are shown as blue dots while WDs with MS companions are shown as red dots. The dashed line is a fitted quadratic relation with parameters given in Eq. \eqref{eq:t_Mv_relation}.
\label{fig:Age}}
\end{figure}

The relation between the luminosity and age allows an approximate theoretical estimate of the expected WD luminosity function.  Assuming that the WD formation rate $\dot n_{\rm WD}$ in the last $\sim 3 \,{\rm Gyrs}$ is roughly constant, the expected number of WDs within each luminosity bin $[M_{V,1}, M_{V,2}]$ is given by 
\begin{equation}
N\approx\dot n_{\rm WD}[t_{\rm cool}(M_{V,2})- t_{\rm cool}(M_{V,1})],
\end{equation}
where $t_{\rm cool}(M_V)$ is the cooling age for a WD with magnitude $M_V$ and is approximated by 
the following relation
\begin{equation}\label{eq:t_Mv_relation}
\log_{10}(t_{\rm cool}/{\rm yr})=-0.04 M_V^2+1.46 M_V-3.22,
\end{equation}
which fits the data shown in Figure \ref{fig:Age} and is plotted as a black dashed line.
We find that this expected luminosity function, shown as a black dashed line in Figure \ref{fig:Mv}, matches reasonably well with that of the single WDs for a fitted value of $\dot n_{\rm WD}=0.7\times10^{-12}~{\rm pc}^{-3}{\rm yr}^{-1}$, which is consistent with the rate inferred by \citet{Liebert05}. The expected luminosity function is calculated for $M_V<14.5$, roughly corresponding to an age of $3\rm \,Gyr$. At significantly older ages, the assumption of constant formation rate is not justified. While the shape of the expected luminosity function is approximately consistent with that of the single WDs, which are dominated by faint WDs, it is inconsistent with that of the WDs with MS companions, for which the luminosity distribution is roughly uniform. If the WDs with MS companions in the \citet{Holberg08} sample distribute according to the above-mentioned luminosity function due to cooling, we expect 1.0 WD detected with $M_V <11.5$ (normalized to the number WDs with $M_V<14.5$ having MS companions) but 5 are in the sample, and there is $0.4\%$ chance that this is due to Poisson fluctuations.

We note that the majority of the MS companions are $M$ and $K$ dwarfs and these systems are unlikely to change as the bright WDs cool. In fact, only one of the $24$ systems has a companion with a main sequence life time shorter than $\sim$Gyr for which near future stellar evolution may affect the binary (Sirius B with an  A0V companion). 

While the small number statistics does not allow a firm conclusion, the shortage of observed faint WDs with MS companions likely indicates a significant incompleteness.  Indeed, a typical WD with $M_V=14$ has a high contrast ($\Delta V\approx 9$) with a Sun-like companion, which makes it very challenging to detect at $\sim 1\arcsec$, corresponding to typical FGK binary separations when placed at $\sim 20\,{\rm pc}$ \citep{Duquennoy91,Raghavan10}. 

How many WDs are missing within $20$ pc?  The amount of bright WDs with MS companions is roughly equal to the amount of single bright WDs. While it is difficult to make accurate quantitative estimates based on the small number statistics, it is likely that the amount of WDs with MS companions is roughly equal or higher than the amount of single WDs. Note that some bright WDs are also likely missed due to very bright companions (some indication for this is the fact that the only two systems with companions earlier than $G$ in this sample are within $5$ pc). If true, at least $\sim 60$ WDs are hiding in binary systems. If the population of single WDs is only $80\%$ complete with about $20$ isolated WDs missing, and if $60\%$ of WDs have a lower mass companion \citep[consistent with the MS observations][]{Raghavan10}, the amount of WDs may be more than double the currently observed one with more than 100 missing WDs. Note that some binaries will evolve to become double WD systems, but this is probably a small fraction given that most companions will not be massive enough to evolve during the lifetime of the Galaxy. 

In contrast to the above discussion, \citet{Holberg08} claimed that the sample of WDs within $20$ pc is $\sim 80\%$ complete (in terms of the total number of WDs). Their main argument was that the distribution of distances is consistent with a uniform distribution in space and that the number density is consistent with the sample within $13$ pc which was claimed to be nearly complete \citep{Holberg02}. The $13$ pc sample was in turn claimed to be complete due to the same space distribution argument along with claims that the apparent brightness of the WDs and the large proper motion are sufficiently large to be detected at $13$ pc. The contribution of unobserved WDs in systems with a MS companion was estimated by these authors to be small based on their small \emph{observed} fraction. None of the arguments presented in  \citet{Holberg08}  and \citet{Holberg02} applies to the likely possibility that a large fraction of WDs have MS companions and are missed due to the high contrast. In fact, the same authors note that the vast majority of systems with MS companions that are of spectral type $K$ or earlier (termed `Sirius-like') are probably missed at distances beyond $20\,\rm pc$ \citep{Holberg09,Holberg13}. In particular they note that there is only one such system in the range of $d=20-25~\rm pc$ which has approximately equal volume as the $d<20~\rm pc$ sample where 11 such systems were found. These authors do not provide an explanation for the sudden suggested drop in completeness at $d=20\,\rm pc$. In fact, $4$ such systems are within $10$ pc, compared to $7$ in the range $10-20$ pc (with a volume which is $7$ times larger than $d<10$ pc) indicating that the completeness fraction is decreasing at increasing distance well within the $20$ pc sample,  providing further evidence for a large incompleteness at $d<20\,\rm pc$. 

\section{Conclusion}
Searches for the missing WDs may resolve the above discrepancy and improve our understanding of the WD population.   
The multiplicity rate of WDs may also have important implications for studying the
progenitor systems of type Ia supernova (SN). In particular, the presence of a MS
tertiary may enhance the rate of WD-WD mergers or collisions -- possible channels for SNe Ia \citep{thompson11, KatzDong12}. In fact, it has been recently suggested that direct WD collisions in triple systems may be the primary channel for SNe Ia \citep{KatzDong12,Kushnir13,Dong14} resolving several key theoretical and observational challenges of these SNe \citep{Kushnir13,Dong14}. WD-WD binaries in quadrupole or higher multiples may be similarly important \citep{Pejcha13}.


\acknowledgments   
We thank Avishay Gal-Yam and Eran Ofek for a discussion that boosted this work and for useful comments. We thank Scott Tremaine for useful comments. S.D. is supported by Òthe Strategic Priority Research Program-The Emergence of Cosmological Structures of the Chinese Academy of Sciences (Grant No. XDB09000000)Ó. D.~K. gratefully acknowledges support from Martin A. and Helen Chooljian Founders' Circle.

\bibliographystyle{apj}

\end{document}